\documentclass[manuscript,format=acmsmall, review=false, screen=true]{acmart}

\usepackage{booktabs} 

\usepackage{todonotes}

\usepackage{tikz}
\usetikzlibrary{arrows,automata,positioning,arrows.meta,shapes.geometric,decorations.pathreplacing}

\usepackage{url}
\usepackage{booktabs}

\usepackage{multirow}
\usepackage{bigdelim}

\usepackage{dataplot}

\usepackage{amsmath,amssymb}
\usepackage{url}
\usepackage{mathdots}

\usepackage{mathtools}

\usepackage{filecontents}

\usepackage{balance}
\usepackage{enumitem}
\usepackage{subfig}

\newcommand{\specialcell}[2][l]{%
  \begin{tabular}[#1]{@{}l@{}}#2\end{tabular}}

\newtheorem{Example}{Example}

\usepackage{etoolbox} 
\makeatletter
\patchcmd{\maketitle}{\@copyrightspace}{}{}{}
\makeatother

\usepackage{tabularx}  

\usepackage{pifont}

\usepackage{graphicx}

\usepackage[ruled]{algorithm2e} 

\SetAlFnt{\small}
\SetAlCapFnt{\small}
\SetAlCapNameFnt{\small}
\SetAlCapHSkip{0pt}
\IncMargin{-\parindent}

\renewcommand\footnotetextcopyrightpermission[1]{} 
\begin{document}
\setlength{\abovedisplayskip}{2pt}
\setlength{\belowdisplayskip}{2pt}
\setlength{\abovedisplayshortskip}{1pt}
\setlength{\belowdisplayshortskip}{1pt}

\title{Advanced Simulation of Droplet Microfluidics}

\author{Andreas Grimmer}
\affiliation{%
  \institution{Johannes Kepler University Linz}
  \country{Austria}}
\email{andreas.grimmer@jku.at}
\author{Medina Hamidovi\'{c}}
\affiliation{%
  \institution{Johannes Kepler University Linz}
  \country{Austria}
}
\author{Werner Haselmayr}
\affiliation{%
 \institution{Johannes Kepler University Linz}
 \country{Austria}}
\author{Robert Wille}
\affiliation{%
  \institution{Johannes Kepler University Linz}
  \country{Austria}
}

\begin{abstract}
The complexity of droplet microfluidics grows with the implementation of
parallel processes and multiple functionalities on a single device.
This 
poses a serious challenge to
the engineer designing the corresponding microfluidic networks.
In today's design processes, the engineer relies on calculations, assumptions, simplifications, as well as his/her experiences and intuitions.
In order to validate the obtained specification of the microfluidic network, 
usually a prototype is fabricated and physical experiments are conducted thus far.
In case 
the design does not implement the desired functionality, this prototyping iteration is repeated -- obviously resulting in an expensive and time-consuming design process.
In order to avoid unnecessary debugging loops involving fabrication and testing, simulation methods could help
to initially validate the specification of the microfluidic network before any prototype is fabricated.
However, state-of-the-art simulation tools come with severe limitations, which
prevent their utilization for practically-relevant applications. More precisely, 
they are often not dedicated to droplet microfluidics, cannot handle the required physical
phenomena, are not publicly available, and can hardly be extended. 
In this work, we present an advanced simulation approach for droplet microfluidics which 
addresses these shortcomings and, eventually, allows to 
simulate practically-relevant applications. 
To this end, we propose a simulation framework which directly works on the specification of the design, supports essential physical phenomena, is publicly available, and easy to extend.
Evaluations and case studies demonstrate the benefits of the proposed simulator: While current \mbox{state-of-the-art} tools were not applicable for practically-relevant microfluidic networks, the proposed solution allows to reduce the design time and costs e.g.~of 
a drug screening device from one person month and USD 1200, respectively, to just a fraction of that.
\end{abstract}

%
%


%
%


\maketitle
\thispagestyle{empty}

\renewcommand{\shortauthors}{Grimmer et al.}

\section{Introduction}\label{s:Intro}
Droplet microfluidics is a highly dynamic and fast evolving field, whose
applications target the fields of chemistry, biology, and material science~\cite{chakrabarty2005design_S}. 
These high dynamics in droplet microfluidics are demonstrated by the rapidly growing number of publications
as demonstrated in~\cite{chou2015recent_S}.
For droplet microfluidics, two different platforms exist: in the
platform which is most frequently used in the  microfluidics domain,
droplets flow through closed microfluidic channels~\cite{teh2008droplet_S}, 
while in the \emph{digital microfluidics} platform the droplets are moved on a hydrophobic surface using 
electrowetting-on-dielectric~\cite{su2008high_S}. 
In this work, we focus on the first platform.

Here, a pump produces a force which causes a flow of a continuous fluid through the 
microfluidic network.
Into this continuous fluid, another immiscible fluid is injected (using e.g.~T- or Y-junctions~\cite{gu2011droplets_S}) which forms droplets. 
Then, the continuous flow transports the droplets through the microfluidic system consisting of channels and modules.
When the microfluidic system consists of multiple paths through which the droplet can flow, 
the resulting designs are called \emph{microfluidic networks}.

The design and realization of microfluidic networks is a complex task which involves the consideration of various aspects such as 
the geometry of the channels, the used phases, the applied pressure, and the effects of droplets.
Engineers consider these aspects by conducting calculations (in the best case, they use custom scripts for this), by trusting their experience, or even rely on their intuition. 
Moreover, in order to tackle the complexity, they frequently apply simplifications and assumptions, e.g.~to ignore the hard to grasp effects of droplets and their corresponding collective hydrodynamic feedback.
Once a complete specification has been derived, a prototype is fabricated next, which is used to evaluate whether 
the resulting design indeed implements the desired functionality or not.
If this is not the case (which is likely in the first iterations), the engineer has to refine the design and continue the entire process again -- including another fabrication and evaluation.
This prototyping cycle is repeated until the engineer obtains a design which realizes the desired behavior.
As reported e.g.~in works such~\cite{chen2017microfluidic_S,erickson2005towards_S}, this can be a difficult, time-consuming,
and expensive process.

In order to address this problem, simulation approaches utilizing the \emph{one-dimensional (1D) analysis model}
have been proposed (see e.g.~\cite{glawdel2011passive_S,jousse2006bifurcation_S,smith2010agent_S,oh2012design_S,sessoms2009droplet_S,sessoms2010complex_S,behzad2010simulation_S,cybulski2010dynamic_S,schindler2008droplet_S,biral2015modeling_S,gleichmann2015toward_S}), which aim to allow for an evaluation of the specification of the design prior to fabrication\footnote{Note that, 
also simulation approaches utilizing Computational Fluid Dynamics (CFD) are available.
They, however, focus more on details such as droplet deformation and splitting and are usually
not capable to simulate entire microfluidic networks. This is discussed later in Section~\ref{s:relatedWork}.}.
More precisely, they allow to predict
(a)~the droplets' paths/trajectories through the network
(this can decide which assay is executed on the droplet),
(b)~the flow changes caused by all droplets and the resulting impacts (e.g.~distance changes between droplets, droplet patterns, etc.), and
(c)~the time a droplet takes to pass through the network.
Using these functionalities, the design could initially be validated before the first prototype is fabricated, 
alternative designs could be explored (i.e.~to test different dimensions of channels, applied pressures, etc.), and 
the design could be optimized.

However, none of the existing solutions got established in practice yet. 
This is caused by the fact that, despite their promises, currently available simulation solutions 
\begin{enumerate}
\item are not dedicated to microfluidics and, therefore, first require to manually map the design 
  to an electrical circuit 
or can only determine a single, static state of the microfluidic network, but do not allow 
to simulate the time-dynamic behavior caused by the flow of droplets
  as it is the case for Spice~\cite{oh2012design_S},
\item target only networks which consist of channels which branch and merge~\cite{schindler2008droplet_S,biral2015modeling_S,gleichmann2015toward_S} or are even limited to networks consisting of a symmetric/asymmetric loop~\cite{glawdel2011passive_S,jousse2006bifurcation_S,smith2010agent_S,sessoms2009droplet_S,sessoms2010complex_S,behzad2010simulation_S,cybulski2010dynamic_S},
but ignore essential physical phenomena such as trapping droplets, checking whether droplets are squeezed through any gaps, and clogging of channels,
\item are not publicly available (in fact, no tool is publicly available), and
\item are static, i.e.~do not allow for further extensions which is essential in order to support the broad range of application scenarios engineers in the microfluidic domain are faced with.
\end{enumerate}
As a consequence, the design of microfluidic devices still follows the costly and time-consuming ``trial-and-error'' approach reviewed above.

In this work, we are introducing an advanced simulation approach which addresses these shortcomings. To this end, we propose a 
simulation framework which (1)~directly works on the specification of the design and considers
the interdependencies caused by all droplets, 
(2)~extends the current state of the art with important physical phenomena which are required for practical designs,
(3)~is publicly available at \url{http://iic.jku.at/eda/research/microfluidics_simulation/}, and,
(4)~due to the availability of the source code and the event-based algorithm,
can easily be extended to support further applications.

In the following, the proposed simulation approach is introduced and demonstrated as follows:
Section~\ref{s:Framework} provides an overview of the simulation framework and describes the respectively applied 1D~analysis model it is based on. 
In Section~\ref{s:Advanced}, we consider physical phenomena which cannot be simulated by 
the previously proposed approaches covered above and discuss how support for them can easily be added to the proposed framework. 
In Section~\ref{s:Eval}, we evaluate how the simulation framework advances the state of the art 
(where no simulations at all are applied) and
demonstrate by means of a case study the application of the proposed framework for the design of a \mbox{practically-relevant} microfluidic network.
More precisely, we show that using the proposed framework allows to reduce the design time and costs e.g.~of 
the drug screening device proposed in~\cite{chen2017microfluidic_S} from one person month and USD~1200, respectively, to just a fraction of that.
Finally, we compare the 
proposed simulation framework to related work and especially to simulations
on other abstraction levels (e.g.~\mbox{CFD-simulations}) in Section~\ref{s:relatedWork} and 
conclude the paper in Section~\ref{s:Conclusion}.

%
%
%


\section{Simulation Framework}\label{s:Framework}
This section introduces the main working principle of the proposed simulation framework for droplet microfluidics.
The framework is based on the \emph{one-dimensional (1D) analysis model} as described in~\cite{schindler2008droplet_S}, which
reduces the microfluidic network (i.e.~an object in the 3D-space) to the 1D-space.
In the following, we describe the general idea of the approach, i.e.~how this abstraction is utilized for 
a fast (i.e.~computationally inexpensive) simulation of 
droplet microfluidic networks. Based on that, the remainder of this work covers how, based on that, further physical
phenomena can be added to make the simulation framework applicable for more practically-relevant microfluidic networks.

The framework describes the microfluidic network as a directed graph consisting of \emph{nodes} and \emph{edges}.
The edges represent channels, modules, and pumps. Their direction represents the counting direction of the flow.
The nodes connect the edges to each other.

The \emph{flow state}
of all the channels and the modules within such a network is then described by the Hagen-Poiseuille equation~\cite{bruus2008theoretical}, i.e.~by
$\Delta P = R\,Q$, where $\Delta P$ is 
the \emph{pressure difference} (in~$[mbar]$) between the two end nodes of the channel/module, 
$Q$ is the \emph{volumetric flow rate} (in $[\mu l/min]$) through the channel/module,
and $R$ is the \emph{fluidic resistance} (in $[mbar/(\mu l/min)]$) posed by the channel/module.
A low Reynolds number allows to reduce the resistance of channels/modules
(which is defined by their geometry and the viscosity of the continuous phase~$\mu_{cont}$)  
to a constant value (i.e.~a reduction from the 3D-space into the 1D-space).
For example, the resistance~$R_c$ of a rectangular channel~$c$ (with length~$l_c$, width~$w_c$, and height~$h_c$),
where the ratio $h_c/w_c$ is less than~$1$, is defined by~\cite{fuerstman2007pressure_S}
\begin{align}\label{eq:resistance}
  R_c = \frac{a\,\mu_{cont}\,l_c}{w_c\,h_c^3},
\end{align}
where $a$ denotes a dimensionless parameter defined as
\begin{align}
  a = 12 \left[ 1 - \frac{192\,h_c}{\pi^5\,w_c} \text{tanh}\left(\frac{\pi\,w_c}{2\,h_c}\right) \right]^{-1}.
\end{align}

Also the pumps producing the flow through the microfluidic networks can be described in the 1D-space:
A syringe pump produces a constant volumetric flow rate~$Q_{in}$ and a 
peristaltic pump produces a pressure gradient~$\Delta P_{in}$.

The presence of droplets in channels/modules change the flow state as they cause additional 
resistances.
Current state-of-the-art simulation tools track the droplets 
as infinitely small points in the channel/module (later in Section~\ref{ss:Implementation} a more comprehensive model for droplets
is introduced and applied)
and a sufficient large distance between droplets (typically a few channel sections/diameters) prevent that
their flow perturbations interact~\cite{schindler2008droplet_S}. These assumptions allow to model each droplet 
by an additional resistance, which is again a value in the 1D-space. 
When $n$ droplets flow through a channel/module,
the overall resistance can be calculated by 
\begin{align}
R^\star = R_c + n\,R_{d}.
\end{align}

The droplet resistance~$R_d$ has been experimentally studied in several works as e.g.~\cite{glawdel2012global_S,fuerstman2007pressure_S,biral2013introducing_S}.
For example, \cite{glawdel2012global_S} established the rule that each droplet
increases the resistance of the segment of channel it occupies by 2-5 times. 
When using a factor of~3, the droplet
resistance is described by 
\begin{align}
R_d = \frac{3\,a\,\mu_{cont} L_d}{w_c\,h_c^3}, 
\end{align}
where $L_d$ is the droplet length.

In order to determine the flow states in all edges, the framework 
automatically applies the mass conservation at each node of the graph
and the relation described by the Hagen-Poiseuille equation~\cite{schindler2008droplet_S}.
The obtained equations are similar to the well-known Kirchhoff's law and
can be directly transferred when we map the Hagen-Poiseuille equation
to the \emph{Ohm's law} with $V = R\,I$ (where the voltage~$V$ corresponds to the pressure gradient~$\Delta P$,
the current~$I$ corresponds to the volumetric flow rate~$Q$, and the resistance~$R$ of a conductor corresponds to the fluidic
resistance~$R$).
More precisely,
\vspace{-0.07cm}\begin{itemize}
\item the sum of flow rates into a node is equal to the sum of flow rates out of that node and
\item the directed sum of pressure gradients around any closed cycle in the graph is zero.
\end{itemize}\vspace{-0.07cm}

By solving the obtained equation system, the framework derives the flow state (i.e.~$\Delta P$ and $Q$) in every channel and module
for the \emph{current} droplet positions. 
The obtained flow rates in the channels/modules determine the \emph{current}
speed of the droplet by $v_d = \alpha \cdot Q / (w_c h_c)$, where $\alpha$ is the slip factor.
Under the conditions where the droplet length is between $1.5$ and $7.2\cdot w_c$, the
viscosity ratios $0.03$ or $0.88$, and the capillary number between $0.001$ and $0.01$ without surfactant,
Vanapalli et al.~\cite{vanapalli2009hydrodynamic_S} found the slip factor to be constant and equal to $\alpha =1.28$.

Using this model, the framework can predict the traffic of droplets, i.e.~when
a droplet arrives at a bifurcation
it chooses the branch with the instantaneous highest flow rate~\cite{engl2005droplet_S,jousse2006bifurcation_S,glawdel2011passive_S}
and does not split (this is true at a low capillary number because the surface tension dominates the viscous stress).

The introduced equations allow to determine the flow state as well as droplet velocities for a certain droplet state
and, hence, update the system state including the droplet positions.
The respectively obtained flow state is valid until 
\vspace{-0.07cm}\begin{itemize}
\item a new droplet is injected (adds a resistance),
\item any droplet leaves the network (removes a resistance), or
\item any droplet enters another edge (causes a shift of the resistance).
\end{itemize}\vspace{-0.07cm}
Hence, as soon as any of those \emph{events} occurs, the current flow state becomes invalid and the simulation
framework re-calculates the flow state (i.e.~newly added, removed, or changed resistances are
incorporated into the equation system which, afterwards, is re-solved).
These event-based calculations make the framework suitable to efficiently simulate large microfluidic networks.

Overall, this provides the basic principle of an efficient simulation of droplet-based microfluidic networks 
(abstracting the channels and droplets to 1D values, use the Hagen-Poiseuille and mass conversation laws to determine
the flow state in all channels/modules, update the droplet positions, 
and adjust/re-evaluate the equation system for the next event). 
However, this state of the art simulation approach
only provides the basics for simulating droplets flowing through networks consisting of channels which
branch and merge (exploiting the fact that a droplet always flows into the branch with
the highest instantaneous flow rate). 
But it does not yet support the simulation of droplets on which actual operations are executed.
How this framework can be extended with the correspondingly required physical phenomena is covered in 
the next section.


\section{Advanced Simulation Framework}\label{s:Advanced}
This section describes how more advanced physical phenomena can be simulated using the proposed framework. 
We are illustrating that by representative (and practically-relevant) examples from the literature, i.e.~proposed designs introduced in
the recent past which have been designed without simulation support (as current simulators were not suited).
To this end, we first review those designs and what phenomena were missing in existing simulation approaches to properly
simulate them.
Afterwards, we describe how support for those phenomena are integrated into
the framework eventually allowing for simulating these designs.

\begin{figure}
  \centering
	\subfloat[Realization of passive trapping wells~\cite{chen2017microfluidic_S}\label{fig:trappingWell}]{
	\includegraphics[width=0.5\columnwidth]{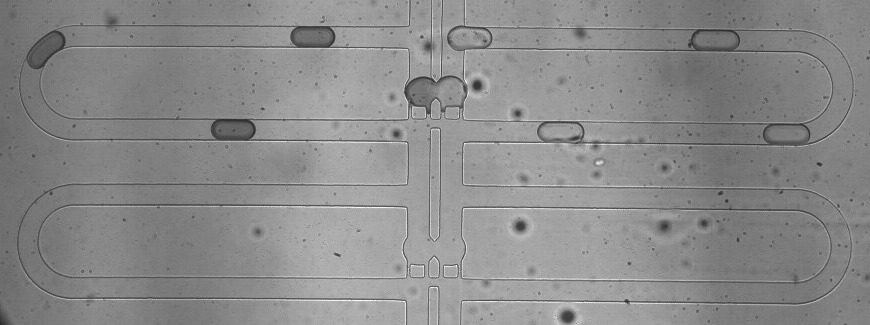}
	}
	\subfloat[Design of a multi-drop switch~\cite{castorina2017microfluidic_S}\label{fig:multiDropSwitch}]{
	\includegraphics[width=0.5\columnwidth]{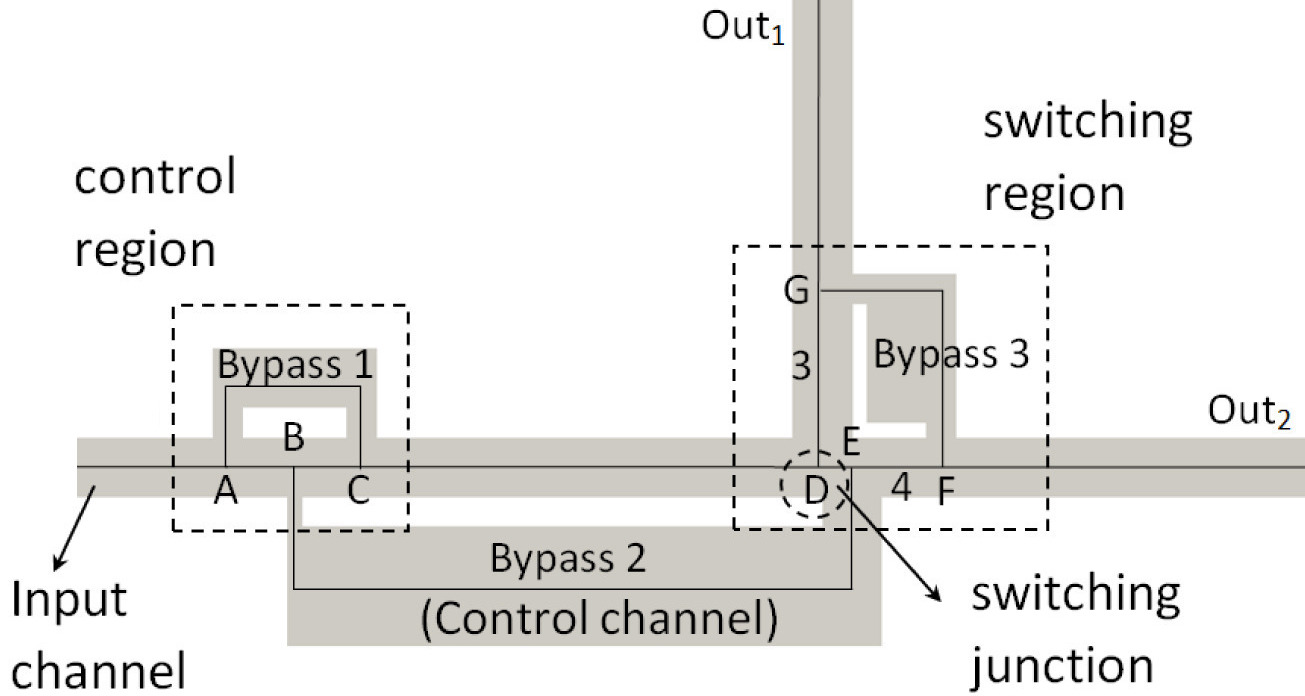}
	}  
  \vspace{-0.3cm}\caption{Physical designs which have been designed without simulation support}\label{fig:designs}\vspace{-0.3cm}
\end{figure}

\subsection{Unsupported Phenomena}\label{ss:Networks}
An important operation in experiments is to \emph{trap} droplets as
these trapped droplets allow to precisely control the reaction time or to observe particle-particle interactions~\cite{wang2009demand_S,chen2017microfluidic_S}.
Figure~\ref{fig:trappingWell} shows a passive realization (i.e.~only hydrodynamic effects and no external 
components for control are used) of two connected trapping well pairs,
which have been proposed in~\cite{chen2017microfluidic_S}.
If a trapping well does not yet contain a droplet (which is the case for the second trapping well pair),
the first arriving droplet is trapped. As soon as it contains a droplet,
all following droplets do not enter the trapping well anymore and are bypassed.
Therefore, the design ensures that the trapped droplet clogs the 
two narrow channels so that
the flow into the bypass channel is higher.
Additionally, in order to analyze the trapped droplet, it must not be squeezed through the narrow channels.

Besides trapping wells, switches are essential in many applications in order to control the path of droplets.
In~\cite{castorina2017microfluidic_S}, the switch shown in Figure~\ref{fig:multiDropSwitch} has been recently proposed,
which is capable to route multi-droplet frames.
This switch uses the effect that the presence of a droplet at the input of a narrow channel causes a blocking of the flow 
into this channel (i.e.~the droplet \emph{clogs} the channel).
If a droplet flows through the ``control region'' shown in Figure~\ref{fig:multiDropSwitch}, 
it clogs the flow into the channel downwards named ``Bypass~2''.
This flow change caused by the clogging is used to route other droplets in the ``switching region'' 
to one of the outputs named ``Out$_1$'' and ``Out$_2$''.

However, 
in order to simulate these operations, the 
simulation of further physical phenomena is required, namely whether droplets 
\vspace{-0.07cm}\begin{itemize}
\item are \emph{trapped} in the microfluidic network,
\item are \emph{squeezed} through any gap, or
\item are \emph{clogging} a channel.
\end{itemize}\vspace{-0.07cm}
Using the ``basic'' framework introduced in the previous section, none of these phenomena and, hence, 
none of the operations can properly be simulated. Since also all related work proposed thus far does not provide support for that,
the practically-relevant applications discussed in~\cite{chen2017microfluidic_S,castorina2017microfluidic_S}
cannot be simulated thus far.

\subsection{Implementation of the Phenomena}\label{ss:Implementation}
\begin{figure}
\centering
\begin{minipage}{.4\textwidth}
  \centering
  \includegraphics[scale=0.16]{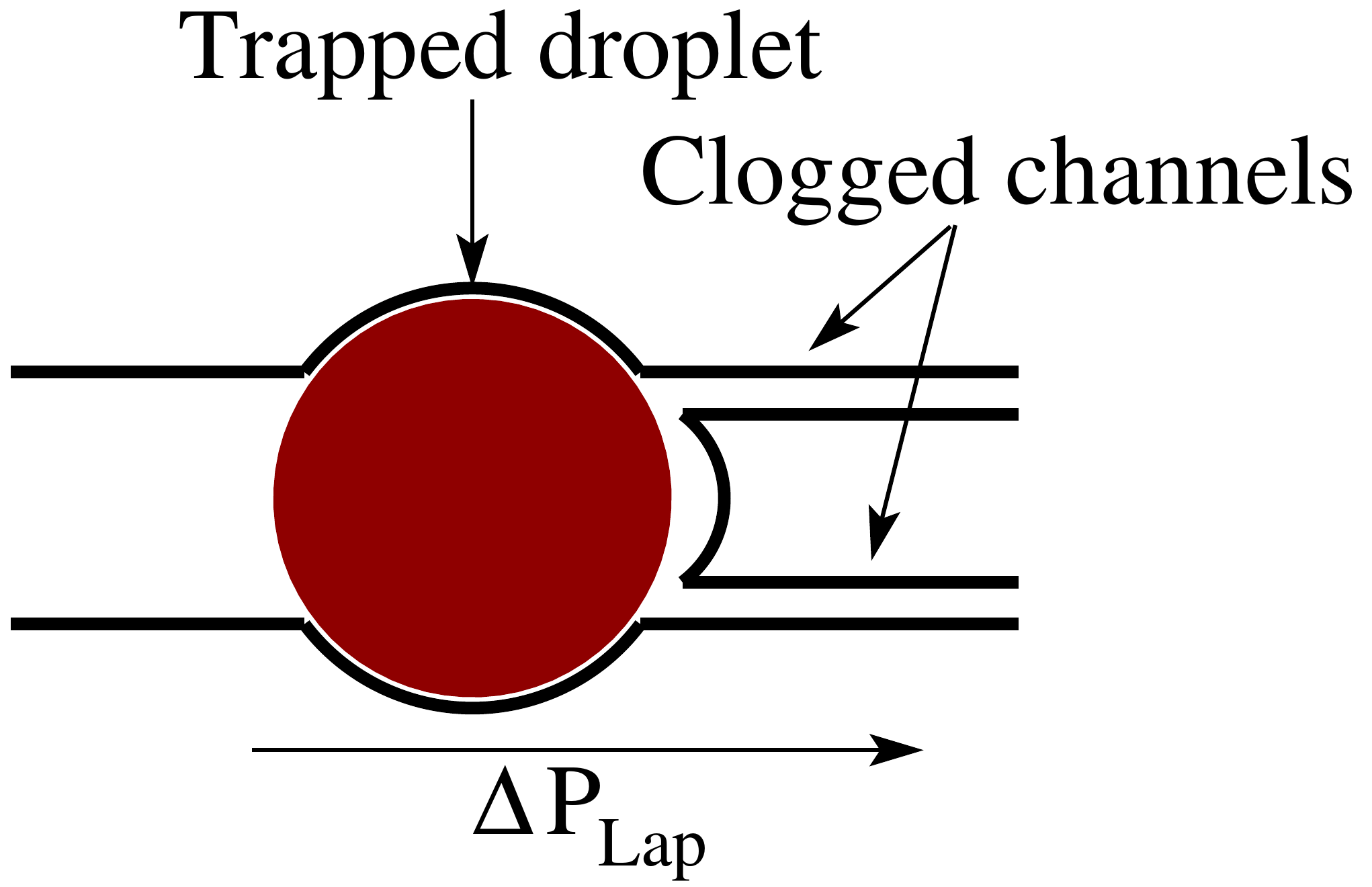}
  \captionof{figure}{Schematic of a trapped droplet}
  \label{fig:trappedDropletSchematic}
\end{minipage}%
\begin{minipage}{.6\textwidth}
	\centering
	\subfloat[Start of clogging\label{fig:startClogging}]{
	\includegraphics[scale=0.12]{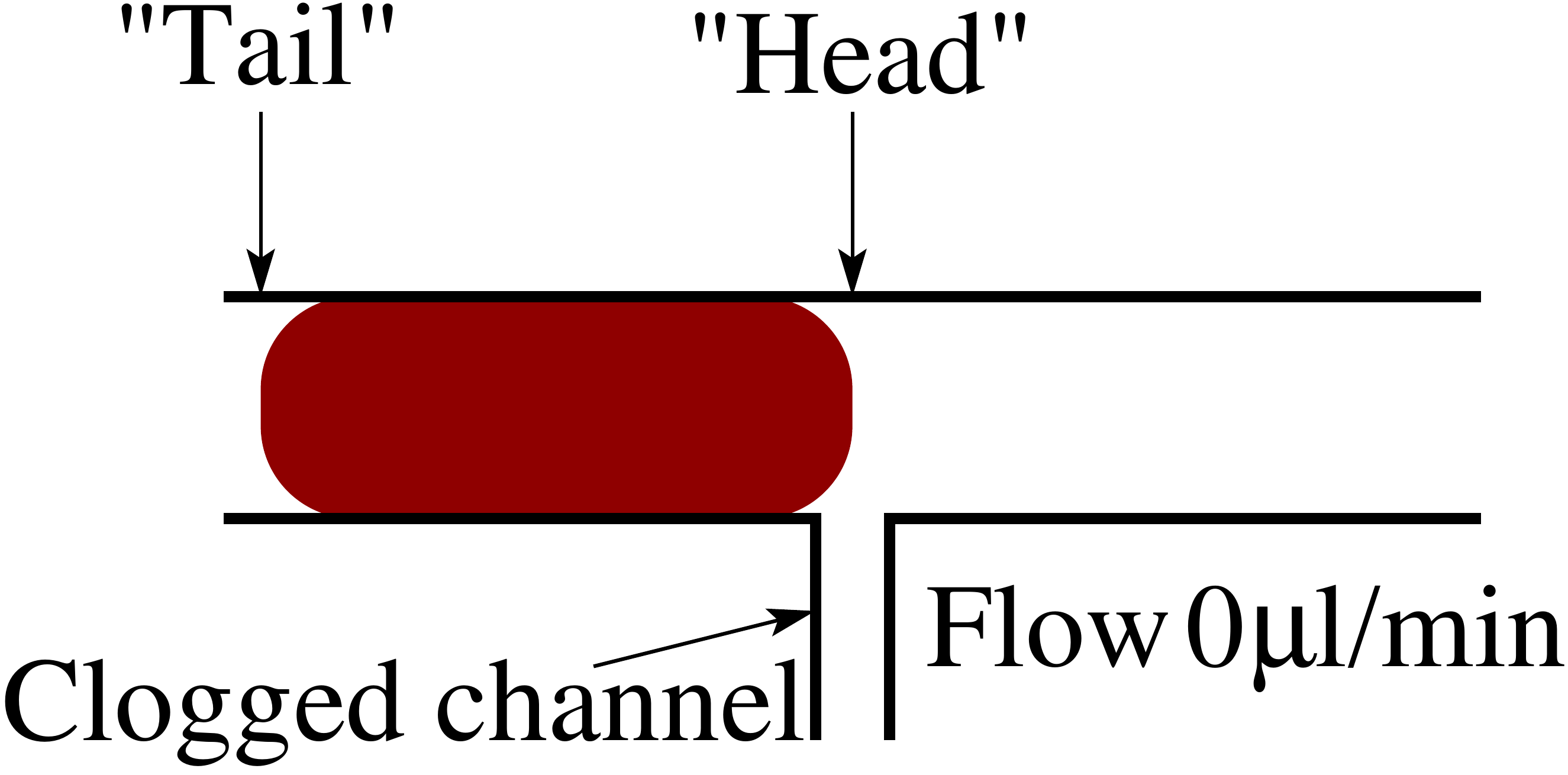}
	}
	\subfloat[End of clogging\label{fig:endClogging}]{
	\includegraphics[scale=0.12]{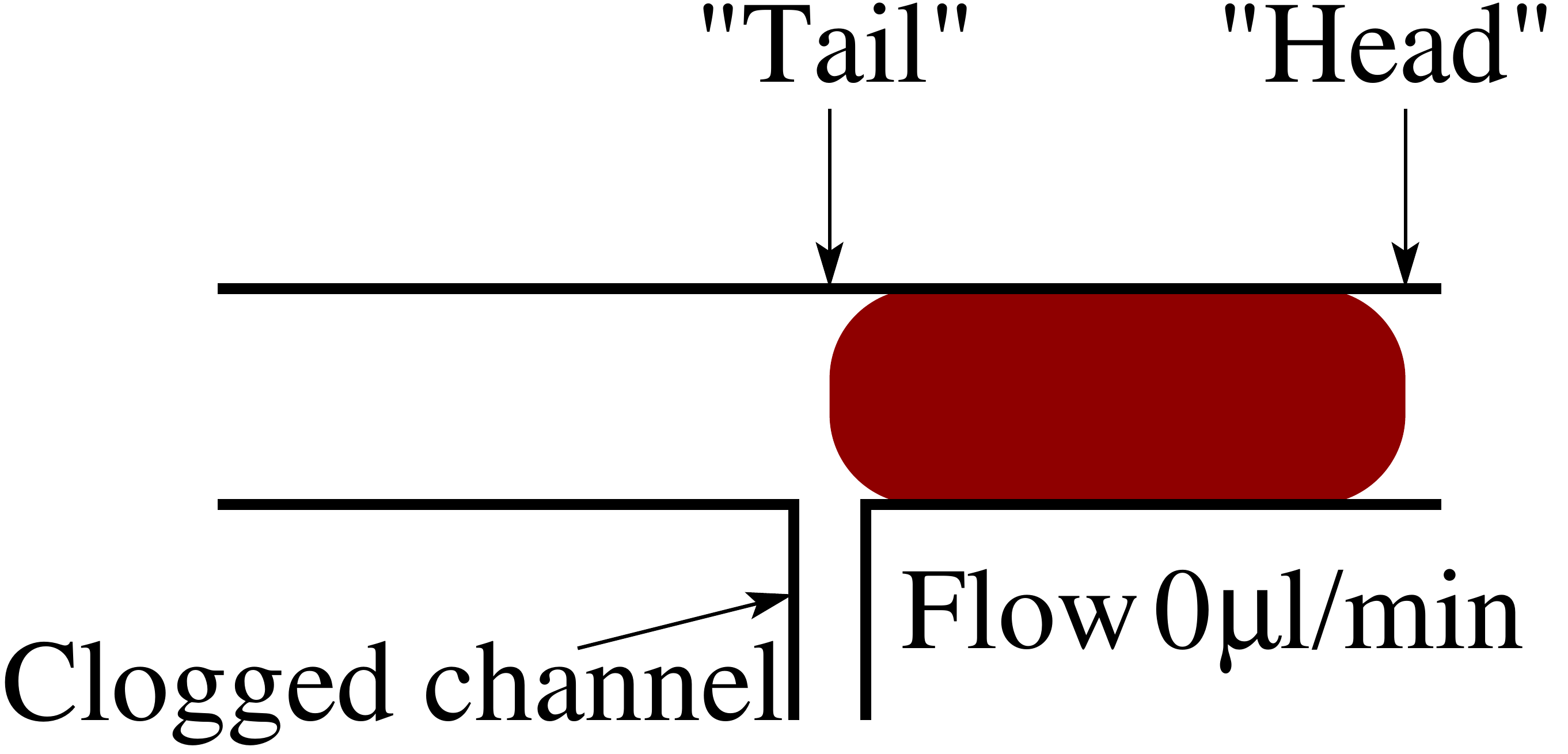}
	}
	\caption{Clogging time span}
	\label{fig:Clogging}
\end{minipage}\vspace{-0.5cm}
\end{figure}

In order to support these physical phenomena and, by this, allow for the simulation of practically-relevant applications such as those discussed above, we extend the 
introduced framework with new equations and events.
These new events demonstrate how the presented framework allows for  easy extensions --
here in form of the following three events:

\textbf{Droplet Trapped Event:}
A droplet is trapped in the microfluidic network, when it stops in an edge.
As long as a trapped droplet is not pushed further (e.g.~by a change of the pressure), it stays in the edge (potentially until 
the end of the simulation).
In the framework, this event is triggered when a droplet is contained in an edge (i.e.~a channel or module),
which does not have a successor edge through which the droplet can leave this edge (cf.~the next event implements
the check whether a droplet is pushed out of an edge).
\vspace{-0.12cm}\begin{Example}
Figure~\ref{fig:trappedDropletSchematic} shows a schematic of a trapping well
with two narrow successor channels (i.e.~having small widths), which prevent the trapped droplet to enter.
When a droplet is fully contained in the trapping well, the respective event is triggered.
\end{Example}\vspace{-0.12cm}


\textbf{Droplet Squeezed Through Gap Event:}
A droplet is squeezed through a gap, when the \emph{Young-Laplace} equation is not fulfilled.
This Young-Laplace equation determines
whether the resulting pressure deforms the droplet and squeezes the droplet through the gap.
More precisely, the Young-Laplace equation is defined as
\begin{align}
\Delta P_{Lap} < \gamma \left[\left(\frac{2}{w_{gap}} + \frac{2}{h}\right) - \left(\frac{1}{r_d} + \frac{2}{h}\right) \right],
\end{align}
where $\gamma$ is the interfacial tension (in [$mN/m$]), $w_{gap}$ is the width of the gap, 
and $r_d$ the droplet radius.
That means, when the pressure acting on the droplet 
is smaller than the right term, the droplet is not squeezed 
through any gap (and, therefore, e.g.~stays in the trap).

As the pressure acting on the droplet changes in each system state (i.e.~it depends on all other droplets), 
checking whether a droplet is squeezed through any gap has to be done for all system states.
Therefore, the simulation framework is extended so that 
every time a new system state is determined, the obtained pressures are checked whether they 
exceed the Young-Laplace pressures. If so, a corresponding event is triggered.

\vspace{-0.1cm}\begin{Example}
Consider again the example shown in Figure~\ref{fig:trappedDropletSchematic}.
When the pressure~$\Delta P_{Lap}$ acting on the droplet is smaller than the \mbox{Young-Laplace} pressure, 
the droplet stays in the trap. 
As this pressure~$\Delta P_{Lap}$ is part of the current system state, it has to be checked for all system states.
In case the pressure exceeds the Young-Laplace pressure, the simulation framework reports this to the engineer and terminates
the simulation.
\end{Example}\vspace{-0.1cm}

\textbf{Droplet Starts/Ends Clogging Event:}
A droplet clogs the flow into an edge when it blocks the input of this edge but does not enter this edge.
In both of the 
microfluidic networks discussed in Section~\ref{ss:Networks}, droplets are used to clog the flow:
In the network proposed in~\cite{chen2017microfluidic_S}, a trapped droplet is pushed 
by the pressure against two narrow gaps and, hence, clogs the flow into these narrow gaps
(cf.~the trapping well in Figure~\ref{fig:trappedDropletSchematic}).
In the switch proposed in~\cite{castorina2017microfluidic_S}, the flow 
into the perpendicular channel (cf.~the channel arrangement in Figure~\ref{fig:Clogging}) is clogged
when a droplet passes.

However, the geometric information which is required to decide
whether a droplet can clog a channel 
is not available in the applied model as it abstracts the 3D-network to 1D-values. 
Therefore, we extend the simulator with a new edge type, i.e.~with \emph{cloggable edges}.
These cloggable edges allow to model that a passing or trapped droplet blocks the flow into this edge.
More precisely, the flow into a cloggable edge is blocked in the following two cases:
First, when a cloggable edge and an edge containing a trapped droplet are connected to the same node 
(i.e.~the trapped droplet clogs the flow, cf.~the trapping well).
Second, when a cloggable edge is connected to a node through which a droplet passes 
(i.e.~the droplet temporary clogs the flow, cf.~the switch).

In order to implement this clogging in the simulation framework, information about the time span
when the droplet clogs the channel is required.
However, this information is not yet available in the framework as presented in Section~\ref{s:Framework}
because the underlying model tracks the droplets as infinitely small points.
This is a disadvantage of this model and, therefore, also of state-of-the-art simulation tools, which
limits the practicality of the state of the art.

In order to allow clogging in the proposed simulation framework, we
extend the model with position information of droplets, 
i.e.~the framework tracks the position of the ``head'' and the ``tail'' of the droplet.
More precisely, this additional position information allows to extend the framework with two new events which are
triggered when a droplet starts or stops clogging a channel.

\vspace{-0.1cm}\begin{Example}
Figure~\ref{fig:Clogging} shows two states of a droplet flowing along a channel. During these 
two states, the narrow channel is clogged by the droplet and, therefore,
the flow into this channel is blocked.
Here, the framework first triggers an event when the ``head'' of the droplet
is located over the narrow channel, which starts the clogging.
Later, when the ``tail'' of the droplet is over the narrow channel,
the framework triggers another event which stops the clogging.
For these two events, the enriched model containing the position information of droplets is used.
\end{Example}\vspace{-0.1cm}

These two events give the time span when a droplet clogs a channel.
In order to implement the blocking of the flow into the clogged edge,
the underlying graph representing the microfluidic network
needs to be dynamically changed.
More precisely, when an event is triggered to start the clogging, the respective edge is removed from the graph.
Similarly, when an event is triggered to stop the clogging, the respective edge is again added to the graph.
These dynamic changes require a re-analysis of the underlying graph,
the derivation of a new equation system, and the re-calculation of the flow states.

\section{Evaluation and Case Study}\label{s:Eval}

The simulation framework proposed above as well as a corresponding 
graphical user interface 
has been implemented in Java (which 
makes the framework platform-independent) and made publicly available
at \url{http://iic.jku.at/eda/research/microfluidics_simulation/}.
The resulting tool addresses the main shortcomings of the current state of the art 
(reviewed in Section~\ref{s:Intro}) e.g.~by being
\vspace{-0.1cm}\begin{itemize}
\item dedicated to droplet microfluidics (it allows to simulate the 
time-dynamic behavior caused by the flow of droplets as well as
it directly uses the specification of the microfluidic network, which is 
both unsupported by Spice),
\item applicable for practically-relevant networks as the framework now supports important physical phenomena,
\item publicly available, and 
\item easily accessible and extendible.
\end{itemize}\vspace{-0.1cm}
By this, the resulting framework has the potential
to establish simulation in the design of droplet microfluidics -- eventually allowing to avoid
unnecessary ``trial-and-error'' iterations with costly and time-consuming physical fabrications.

In order to confirm that, intensive evaluations have been conducted, which 
consist of tests for the unsupported phenomena and the application of the framework for a
case study using the microfluidic network of~\cite{chen2017microfluidic_S}.
In the following, we summarize the most important evaluations.



\begin{figure*}
\subfloat[$t=0\,ms$]{
\includegraphics[scale=0.3]{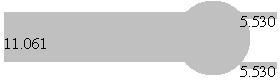}
}\vspace{0.14cm}
\subfloat[$t=6\,ms$]{
\includegraphics[scale=0.3]{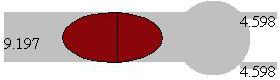}
}\vspace{0.14cm}
\subfloat[$t=12\,ms$]{
\includegraphics[scale=0.3]{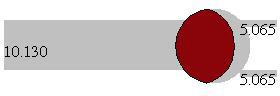}
}\vspace{0.14cm}
\subfloat[$t=14\,ms$]{
\includegraphics[scale=0.3]{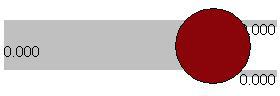}
}
\vspace{-0.7cm}\caption{The framework simulates a trapped droplet, checks the Young-Laplace pressure, and clogs the flow into the gaps.}
\label{fig:trapSnaps}\vspace{-0.6cm}
\end{figure*}

\begin{figure*}
\subfloat[$t=0\,ms$]{
\includegraphics[scale=0.3]{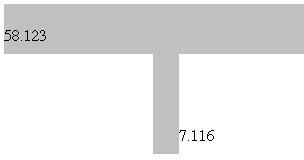}
}
\subfloat[$t=4\,ms$]{
\includegraphics[scale=0.3]{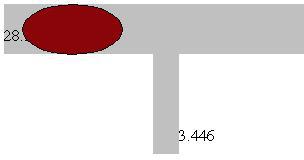}
}
\subfloat[$t=8\,ms$]{
\includegraphics[scale=0.3]{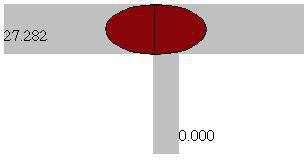}
}
\subfloat[$t=12\,ms$]{
\includegraphics[scale=0.3]{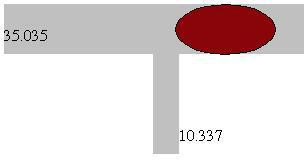}
}
\vspace{-0.3cm}\caption{The framework simulates the clogging using the ``head'' and ``tail'' position information of the droplet.}
\label{fig:switchSnaps}\vspace{-0.5cm}
\end{figure*}

\subsection{Evaluation of the Phenomena}\vspace{-0.1cm}
A main characteristic of the proposed simulation framework 
(which is essential to make the approach broadly applicable for \mbox{practically-relevant} applications as discussed in Section~\ref{ss:Networks})
is its direct support of several physical phenomena which have not been supported yet. 
To demonstrate the working principle of the proposed simulator,
we set up small networks to be simulated which require the corresponding features.
More precisely, we consider
(1)~a network composed of a channel connected to a single trapping well and (2)~a network composed of a channel to which a perpendicular
channel is connected. For both networks, we used similar specifications, which are summarized in 
Table~\ref{tab:specification}.

\begin{table}
\small
\setlength{\tabcolsep}{0.5mm}
\parbox{.45\linewidth}{
\begin{tabular}{l|r} \toprule
Channel height                 & $53\mu m$     \\ \hline
Channel width                  & $100\mu m$    \\ \hline
Trapping well radius           & $75\mu m$     \\ \hline
Gap width                      & $15\mu m$     \\ \hline
Perpendicular channel width    & $30\mu m$     \\ \hline
Applied pressure               & $30mBar$      \\ \hline
Continuous phase (silicone oil) & $4.565 mPa\,s$ \\ \hline
Dispersed phase (water)        & $1 mP\,s$      \\ \hline
Interfacial tension            & $42 mN/m$    \\ \bottomrule
\end{tabular}
\caption{Specification}\vspace{-0.3cm}
\label{tab:specification}
}\hspace{0.1cm}
\parbox{.49\linewidth}{
\begin{tabular}{ccc|l} \toprule
ID & $L_{bypass}$ & $w_{gap}$ & Simulation results\\ \midrule
1  & $3000\mu m$ & $15\mu m$ &  \specialcell{No reliable trapping}\\  \hline
2  &$4000\mu m$ & $15\mu m$ &  --\\ \hline
3  &$5000\mu m$ & $15\mu m$ &  Bypass length decreases throughput\\ \hline
4  &$3000\mu m$ & $25\mu m$ &  \specialcell{Sensitive to high input pressures}\\ \hline
5  &$4000\mu m$ & $25\mu m$ &  \specialcell{Sensitive to high input pressures}\\ \hline
6  &$5000\mu m$ & $25\mu m$ &  \specialcell{Sensitive to high input pressures,\\bypass length decreases throughput}\\ \bottomrule
\end{tabular}
\caption{Reliability}\label{tab:reliability}
}\vspace{-0.8cm}
\end{table}

The accordingly obtained simulation results are respectively provided in
Figure~\ref{fig:trapSnaps} and Figure~\ref{fig:switchSnaps} for selected time steps.
The figures show the determined position of a droplet at each particular time as well as
the instantaneous flow rates (provided in $\mu l/min$) in each channel.

We can observe in Figure~\ref{fig:trapSnaps} that the droplet is successfully trapped in the trapping well. 
Then, the droplet stays in the trapping well, since the Young-Laplace pressure is equal to~$50.4mBar$ (i.e.~as
the droplet entirely fills the trapping well, the trapping well radius is equal to the droplet radius), which
is larger than the applied pressure of~$30mBar$. Hence, the droplet is not squeezed out through any of the two gaps.
Furthermore, as soon as the droplet is fully contained in the trapping well (i.e.~after $14~ms$), 
it blocks the flow into the two narrow channels.

Figure~\ref{fig:switchSnaps} shows that a droplet over a perpendicular channel blocks the flow into this channel.
The simulation framework uses the position of the droplet's ``head'' and ``tail'' in order to determine the time span when 
the droplet clogs the channel.

Overall, these two small networks confirm the correct implementation of the phenomena, 
which is heavily utilized in the following case study. 

\vspace*{-0.2cm}\subsection{Case Study}\label{ss:CaseStudy}\vspace{-0.14cm}
In this case study, we demonstrate the applicability of the proposed simulation framework to a practically-relevant application.
More precisely, 
we consider the design of the microfluidic network proposed in~\cite{chen2017microfluidic_S}.
This microfluidic network is developed to screen drug compounds that inhibit the tau-peptide aggregation, which is a
phenomenon related to neurodegenerative disorders such as Alzheimer's disease~\cite{soto2003unfolding_S}.
For the drug screening, the droplets of different content have to be trapped and merged on demand, which eventually 
allow a precise control of the reaction time.
The working principle is purely passive (i.e.~no valves or other active components are used) and it
is illustrated by means of videos available at \url{https://doi.org/10.1039/C7RA02336G}.
In the following, we first review the design process of this microfluidic network which
has been conducted without any simulations (according to~\cite{chen2017microfluidic_S}).
Afterwards, we show how the proposed simulation framework can help here.

For deriving the specification (i.e.~the channel dimensions, applied pressures, etc.), 
the engineer conducted calculations and applied simplifications as well as assumptions. For example, the engineer
simplified the effects of droplets because it is impossible to consider those by hand.
As the effects of all simplifications and assumptions cannot be assessed,
the engineer came up with six different specifications,
which have three different bypass channel lengths (i.e.~droplets pass this channel when the respective trapping
well already contains a droplet) and two different gap sizes. Table~\ref{tab:reliability} shows the resulting specifications.
In order to validate the functionality of these specifications, the engineer fabricated
prototypes and conducted physical experiments.
In fact, the engineer had no other choice as no simulation tools
were available which would have been capable to handle the required phenomena
(CFD simulations are too computationally expensive for complete microfluidic networks, cf.~Section~\ref{s:relatedWork},
and other state-of-the-art simulations are not applicable).
Finally, the engineer picked the specification trapping the droplets in the most reliable way.
The engineer reported that the fabrication and testing of these six prototypes resulted in one person month of manual labor and
financial costs of USD~1200.

In our case study, we revisited this design process and additionally applied the proposed simulation framework.
Therefore, the different designs were validated and tested
using the framework \emph{before} any physical experiments are conducted.
We set up simulations of the six different specifications (cf.~Table~\ref{tab:reliability}).
Then, we analyzed the obtained simulation results with respect to the taken path of droplets, the flow rates, 
whether droplets are unintentionally squeezed through any gap, and how long it takes until
a droplet is trapped.
For all specifications, we observe the intended paths of the droplets.
But for the specification with ID~1, the simulation shows that the flow rates do not allow a reliable
trapping of droplets (i.e.~the flow into the trapping well was hardly larger than the flow into the bypass).
Next, we simulated different input pressures. Generally, too high input pressures cause
the droplet to be squeezed out of the trapping wells. But we found that especially the specifications with the larger
gap widths (specified with ID~4-6) are more sensitive to higher pressures (i.e.~a larger gap width reduces the Young-Laplace pressure).
Finally, we measured the time until a droplet is trapped as this is especially relevant for bio-assays with cells.
Here, the simulations show that, the longer the bypass channel, the longer the time until a droplet is trapped.
All those effects have also been observed in the physical experiments (but, therefore, six prototypes were necessary).

Table~\ref{tab:reliability} summarizes the obtained results, which 
show a clear preference for the second specification.
This specification is the one which was eventually realized in~\cite{chen2017microfluidic_S}.
For this specification, we provide a video showing the output of the simulator under \url{http://iic.jku.at/eda/research/microfluidics_simulation/}.
Additionally, we compare the simulator's output with photos of a physical experiment captured with a frequency of $50 fps$.
This simulation predicts the same functionality as the corresponding physical experiment, 
which finally allows the utilizing of the simulator to evaluate different specifications. 
Overall, the use of the simulation framework in the design process can reduce 
the number of fabricated designs to a single one, and hence, 
would allow to reduce the design time and costs of this drug screening device from one person month 
and USD~1200, respectively, to just a \mbox{fraction of that}.

\vspace*{-0.22cm}\section{Comparison to Related Work}\label{s:relatedWork}\vspace*{-0.14cm}
In this section, we compare the proposed framework to other simulation tools and levels.
Basically, simulation approaches for droplet microfluidics can be classified into two abstraction levels:
\vspace{-0.1cm}\begin{itemize}
\item \emph{Simulations using Computational Fluid Dynamics (CFD)}: Tools like
\emph{Comsol Multiphysics}, \emph{Ansys}, or \emph{OpenFoam} employ CFD simulations. Comprehensive reviews
of the methods and tools are provided in~\cite{glatzel2008computational_S,worner2012numerical_S}.
These tools simulate the fluid flow in the most accurate way, i.e.~allow to simulate turbulences and effects like droplet 
deformation and splitting.
But therefore, they require a complex simulation 
setup (e.g.~the generation of a mesh based on the physical design).
Furthermore, the high level of physical details causes significant computational costs,
which yield simulation results of high precision 
but also limits their applicability to small designs and single components.
For example, these methods are inappropriate to quickly simulate practically large-scale microfluidic networks~\cite{oh2012design}
and therefore, recently a hybrid solution was presented in~\cite{wang2017instantaneous}, which queries precomputed results from a database 
and combines it with simulations based on the 1D analysis model. 

\item \emph{Simulations on the 1D analysis model:} This model 
is applied in the presented simulation framework and
was introduced in Section~\ref{s:Framework}.
The model is valid when the flow is laminar, viscous, and incompressible~\cite{schindler2008droplet_S}.
The abstraction of the microfluidic network to 1D values makes the simulation efficient as only
linear equations need to be solved, which makes corresponding simulators most suitable for practical large-scale 
microfluidic networks.
These simulations are especially useful for determining the paths and position of droplets, 
the time a droplet takes to pass through the network, as well as for parametric analysis needed
to validate and optimize designs.
However, as discussed in Section~\ref{s:Intro}, existing methods within this category (e.g.~\cite{glawdel2011passive_S,jousse2006bifurcation_S,smith2010agent_S,sessoms2009droplet_S,
sessoms2010complex_S,behzad2010simulation_S,cybulski2010dynamic_S,schindler2008droplet_S,biral2015modeling_S,gleichmann2015toward_S}) 
suffer from limitations such as poor applicability to the microfluidic domain, missing support for  essential physical phenomena, their non-availability, and their rather static and, hence, not extendible nature.
In this work, these shortcomings have been addressed by the  advanced simulation framework.
\end{itemize}\vspace*{-0.1cm}

\vspace*{-0.22cm}\section{Conclusion}\label{s:Conclusion}\vspace*{-0.1cm}
In this paper, we presented an advanced simulation framework which addresses severe limitations of state-of-the-art
simulators by being dedicated to droplet microfluidics and by addressing essential physical phenomena, which are required for practically-relevant applications.
Furthermore, the open-source implementation allows for a broad 
application of the framework and even further extensions.
The resulting framework can be applied by engineers in order to validate their design
before even the first prototype is made.
That simulations can save costs as well as time has been shown in a case study
for a microfluidic network which is used to screening drug compounds 
that inhibit the tau-peptide aggregation, a phenomenon related to neurodegenerative disorders such as Alzheimer's disease.

\bibliographystyle{ACM-Reference-Format} 
\bibliography{../../LoC}

\end{document}